\documentclass[a4]{article}
\usepackage{amsfonts}
\usepackage{amsmath}
\usepackage{amssymb}
\usepackage{graphicx}

\begin{document}

\title{Geodesic motion in the neighbourhood of submanifolds embedded in
warped product spaces}
\author{F\'{a}bio Dahia$^{a}$, Carlos Romero$^{b}$, L\'{u}cio F. P. da Silva,%
$^{b}$ \and Reza Tavakol$^{c}$ \\
$^{a}$Departamento de F\'{\i}sica, Universidade Federal \\
de Campina Grande, 58109-970\\
Campina Grande, Pb, Brazil\\
$^{b}$Departamento de F\'{\i}sica, Universidade Federal da Para\'{\i}ba, \\
Caixa Postal 5008, 58059-979 Jo\~{a}o Pessoa, Pb, Brazil\\
$^{c}$School of Mathematical Sciences , Queen Mary, University of London,\\
London E1 4NS, UK }
\maketitle

\begin{abstract}
We study the classical geodesic motions of nonzero rest mass test particles
and photons in $(3+1+n)$- dimensional warped product spaces. An important
feature of these spaces is that they allow a natural decoupling between the
motions in the $(3+1)$-dimensional spacetime and those in the extra $n$
dimensions. Using this decoupling and employing phase space analysis we
investigate the conditions for confinement of particles and photons to the $%
(3+1)$- spacetime submanifold. In addition to providing information
regarding the motion of photons, we also show that these motions are not
constrained by the value of the extrinsic curvature. We obtain the general
conditions for the confinement of geodesics in the case of pseudo-Riemannian
manifolds as well as establishing the conditions for the stability of such
confinement. These results also generalise a recent result of the authors
concerning the embeddings of hypersurfaces with codimension one.
\end{abstract}


\section{Introduction}

An intriguing idea in modern cosmology is the possibility that the universe
may have a higher number of dimensions than the classically observed $(3+1)$%
. There has been a number of motivations for this idea, mainly related to
attempts at constructing a fundamental theory of physical interactions.
These range from the original attempts at unification of gravity with
electromagnetism by Kaluza and Klein \cite{KK} to String/M-theory \cite%
{Randall,SMS,Maartensrev}.

An immediate task within this higher dimensional framework is how to explain
the four dimensionality of the observed universe. An original idea - which
dates back to the work of Kaluza and Klein - assumes these extra dimensions
to be compact and very small. In most recent braneworld models, on the other
hand, stringy effects are invoked to argue that in low energy regimes
particles are restricted to a special $3+1$ \textit{brane} hypersurface,
which is embedded in a higher-dimensional \textit{bulk}, while the
gravitational field is free to propagate in the bulk \cite{Arkani}.

The intense recent interest in the higher-dimensional scenarios has also
provided strong motivation to look for geometrical mechanisms which could
allow confinement and this has led to the investigation of the so-called
warped product spaces \cite{Frolov} and their geometrical properties.

Originally most braneworld and higher dimensional studies concentrated on $%
(4+1)$ scenarios, i.e. co-dimensional one models. However, given the
prediction of the string theory, according to which spacetime is
10-dimensional, the study of higher co-dimensional models is urgently called
for. Increasing effort has recently gone into the study of such models (see
for e.g. \cite{co-dim}).

An important question concerning such models, specially from an
observational point of view, is the behaviour of the geodesics in these
models, and in particular the relation between the geodesics of the
higher-dimensional space and those belonging to the hypersurface. A great
deal of effort has recently gone into the study of geodesic motions in
five-dimensional spaces \cite{Wesson1}. Recently the authors studied the
ability of five-dimensional warped product spaces to provide a mechanism for
geodesic confinement on co-dimension one hypersurfaces, purely classically
and based on gravitational effects \cite{DRST}. The aim of this article is
to generalise that analysis to the case of co-dimension $n$ warped product
spaces \cite{Bishop}. We do this by studying the classical geodesic motions
of nonzero rest mass test particles and photons in $(3+1+n)$-dimensional
warped product spaces. We show that it is possible to obtain a general
picture of these motions, using the natural decoupling that occurs in such
spaces between the motions in the $n$ extra dimensions and the motion in the 
$4D$ submanifolds. This splitting allows the use of phase space analysis in
order to investigate the possibility of confinement\footnote{%
Throughout by confinement of photons we mean to say that their motion is
constrained to lie in the brane, rather than being bounded.} and the
stability of motion of particles and photons to submanifolds in such $%
(3+1+n) $-dimensional spaces.

The paper is organised as follows. In Section 2 we write the geodesic
equations for warped product spaces and consider the parts due to 4D and the
higher dimensions separately. We then show that the equations that describe
the motion in higher dimensions decouples from the rest. We proceed in
Section 3 to find the mathematical conditions that must be satisfied by the
warping function $f$ in order for timelike and null geodesics of the
higher-dimensional space to be confined to $M^{4}$. In Section 4 we rewrite
the geodesic equations in the higher dimensions as an autonomous $2n$%
-dimensional dynamical system. We then employ phase plane analysis to study
the motion of particles with nonzero rest mass (timelike geodesics) and
photons (null geodesics) respectively. Such qualitative analysis allows a
number of conclusions to be drawn about the possible existence of confined
motions and their nature in the neighbourhood of hypersurfaces. In Section 5
we give an \ analysis of the motion in the extra dimensions by reducing the
problem to the motion of a particle subjected to the action of an effective
potential and illustrate the method with an example from the literature. We
conclude in Section 6 with some final remarks.

\section{Warped product spaces and geodesic motion}

In general a warped product space is defined in the following way. Let $%
(M^{m},h)$ and $(N^{n},k)$ be two\ Riemannian (or pseudo-Riemannian)
manifolds of dimension $m$ and $n$, with metrics $h$ and $k,$ respectively.
Let $f:N^{n}\rightarrow \mathbb{R}$ be a smooth function (which we shall
refer to as the warping function). We can then construct a new warped
product Riemannian (pseudo-Riemannian) space by setting $M=M^{m}\times N^{n}$
and defining a metric $g=e^{2f}h\oplus k$. In this paper we shall take $m=4$
and identify $M^{4}$ with the (3+1)-dimensional \textit{spacetime,}\ a
four-dimensional Lorentzian manifold \ with signature $(+---)$. 
Therefore the class of warped geometries which we shall consider can be
characterised by the following line element 
\begin{equation}
dS^{2}=e^{2f}h_{\alpha \beta }dx^{\alpha }dx^{\beta }-k_{ab}dy^{a}dy^{b},
\label{warpedmetric}
\end{equation}%
where $f=f(y^{1},...,y^{n})$, $h_{\alpha \beta }=$ $h_{\alpha \beta }(x)$
and $k_{ab}$ is in general an $n$-dimensional Riemannian metric\footnote{%
Throughout capital Latin indices take value in the range $(0,1,...(3+n))$,
lower case Latin indices take values in the range $(4,...(3+n))$ while Greek
indices run over $(0,1,2,3).$Thus, the coordinates of a generic point $P$ of
the manifold $M$ will be denoted by $Z^{A}=(x^{\alpha },y^{a})$, where $%
x^{\alpha }$ denotes the 4D spacetime coordinates and $y^{a}$ refers to the $%
n$ extra coordinates of $P$.}.

Let us now consider the equations of geodesics in the $(n+4)$-dimensional
space $M$ 
\begin{equation}
\frac{d^{2}Z^{A}}{d\lambda ^{2}}+\Gamma _{BC}^{A}\frac{dZ^{B}}{d\lambda },%
\frac{dZ^{C}}{d\lambda }=0,  \label{geodesic}
\end{equation}%
where $\lambda $ is an affine parameter and $\Gamma _{bc}^{a}$ are the
Christoffel symbols of the second kind defined by $\Gamma _{BC}^{A}=\frac{1}{%
2}g^{AD}(g_{DB,C}+g_{DC,B}-g_{BC,D})$. After some algebra it is not
difficult to show that the geodesic equations for the 4D part can be written
in the form

\begin{equation}
\frac{d^{2}x^{\mu }}{d\lambda ^{2}}+^{(4)}\Gamma _{\alpha \beta }^{\mu }%
\frac{dx^{\alpha }}{d\lambda }\frac{dx^{\beta }}{d\lambda }=\phi ^{\mu },
\label{4Dequations}
\end{equation}%
where 
\begin{equation}
\phi ^{\mu }=-\Gamma _{ab}^{\mu }\frac{dy^{a}}{d\lambda }\frac{dy^{b}}{%
d\lambda }-2\Gamma _{\alpha b}^{\mu }\frac{dx^{\alpha }}{d\lambda }\frac{%
dy^{b}}{d\lambda },
\end{equation}%
and $^{(4)}\Gamma _{\alpha \beta }^{\mu }=\frac{1}{2}g^{\mu \nu }\left(
g_{\nu \alpha ,\beta }+g_{\nu \beta ,\alpha }-g_{\alpha \beta ,\nu }\right) $%
. Likewise the extra-dimensional part of the geodesic equations (\ref%
{geodesic}) may be written as 
\begin{equation}
\frac{d^{2}y^{a}}{d\lambda ^{2}}+^{(n)}\Gamma _{bc}^{a}\frac{dy^{b}}{%
d\lambda }\frac{dy^{c}}{d\lambda }=\psi ^{a}  \label{extradim}
\end{equation}%
where 
\begin{equation*}
\psi ^{a}=-\Gamma _{\alpha \beta }^{a}\frac{dx^{\alpha }}{d\lambda }\frac{%
dx^{\beta }}{d\lambda }-2\Gamma _{\alpha b}^{a}\frac{dx^{\alpha }}{d\lambda }%
\frac{dy^{b}}{d\lambda }
\end{equation*}

We now assume that our spacetime $M^{4}$ corresponds in this scenario to a \
particular submanifold defined by the $n$ equations $y^{a}=y_{o}^{a}=$%
constant.\ The geometry of $M^{4}$ is then determined by the induced metric 
\begin{equation*}
ds^{2}=g_{\alpha \beta }(x,y_{o}^{1},...,y_{o}^{n})dx^{\alpha }dx^{\beta }.
\end{equation*}
Therefore the quantities $^{(4)}\Gamma _{\alpha \beta }^{\mu }$ which appear
on the left-hand side of Eq.(\ref{4Dequations}) may be identified with the
Christoffel symbols associated with the metric induced on the leaves of the
foliation defined above.

For the warped product space (\ref{warpedmetric}) the quantities $\phi ^{\mu
}$ and $\psi ^{a}$ reduce respectively to 
\begin{equation*}
\phi ^{\mu }=-2f,_{a}\dot{x}^{\mu }\dot{y}^{a},
\end{equation*}%
and 
\begin{equation*}
\psi ^{a}=-f^{,a}e^{2f}h_{\alpha \beta }\dot{x}^{\alpha }\dot{x}^{\beta },
\end{equation*}%
where $f_{,a}=\frac{\partial f}{\partial y^{a}}$, a dot denotes
differentiation with respect to the affine parameter $\lambda $ and we are
using the notation $f^{,a}=k^{ab}f_{,b}$. Note that if the warping function $%
f$ is constant the equations (\ref{geodesic}) which describe the geodesics
of the entire warped product $M$ separates into the geodesic equations for
the two submanifolds $M^{4}$ and $N^{n}$. On the other hand if we restrict
ourselves to the class of timelike or null geodesics of $M$ we have the
following first integral 
\begin{equation}
e^{2f}h_{\alpha \beta }\dot{x}^{\alpha }\dot{x}^{\beta }-k_{ab}\dot{y}^{a}%
\dot{y}^{b}=\epsilon ,  \label{firstintegral}
\end{equation}%
where $\ \epsilon =1$ in the case of particles with nonzero rest mass and $%
\epsilon =0$ in the case of photons. With the help of the above first
integral it is now possible to decouple the motion in the extra dimensions
from the spacetime. Thus the geodesic equations (\ref{geodesic}) finally
take the form 
\begin{equation}
\ddot{}%
x^{\mu }+^{(4)}\Gamma _{\alpha \beta }^{\mu }\dot{x}^{\alpha }\dot{x}^{\beta
}=-2f_{,a}\dot{x}^{\mu }\dot{y}^{a},  \label{spacetime}
\end{equation}%
\begin{equation}
\ddot{y}^{a}+\Gamma _{bc}^{a}\dot{y}^{b}\dot{y}^{c}=-f^{,a}(\epsilon +k_{bc}%
\dot{y}^{b}\dot{y}^{c}).  \label{extradim1}
\end{equation}%
These equations constitute a system of second-order ordinary differential
equations which can, in principle, be solved once the function $%
f=f(y^{1},...,y^{n})$ is given.

\bigskip

\section{Confinement of the motion in four-dimensional spacetime \qquad}

In this section we wish to investigate the possibility of confinement\ of
massive particles and photons in the spacetime manifold $M^{4}$. This
amounts to finding the mathematical conditions required to be satisfied by
the warping function $f$\ such that timelike and null geodesics of $M$\
coincide with those corresponding to $M^{4}$. In general such submanifolds
of a Riemannian\ (or pseudo-Riemannian) manifold $M$ are referred to as
totally geodesic. In the case where $M$ is a Riemannian space, then
according to a theorem of differential geometry the submanifold $M^{4}$ is
totally geodesic if and only if all normal curvatures (or extrinsic
curvatures) of $M^{4}$ vanish \cite{Eisenhart}. If the geometry of $M$ is
pseudo-Riemannian (in our case $M$ is a Lorentzian warped product space)
then, as we shall show below, the theorem is still valid, however the
confinement of null geodesics does not depend on the value of the normal
curvatures of the submanifold $M^{4}$. From a physical standpoint it is
interesting that for the large class of warped product spaces\ defined by
the equation (\ref{warpedmetric}) the motion of photons is not constrained
by the 
extrinsic (or normal) curvature of the spacetime. This result generalises a
recent result of the authors concerning the embeddings of hypersurfaces with
codimension one \cite{DRST}.

To show this, consider equations (\ref{spacetime}) and (\ref{extradim1}).
Let $\gamma $ \ be a timelike (or spacelike) geodesic curve of the
submanifold $M^{4}$. Since $\gamma \in M^{4}$ we must have $\dot{y}^{a}=0$,
which implies that (\ref{spacetime}) is identically satisfied. Now if $%
f^{,a}(y_{o}^{1},...y_{o}^{n})=0$, then (\ref{extradim1}) also holds.
Therefore Eq. (\ref{geodesic}) is satisfied, implying that $\gamma $ is a
geodesic of $M$. Conversely, if any geodesic $\gamma $ of $M$ with
parametric equations $(x^{\alpha }=x^{\alpha }(\lambda ),y^{a}=y_{o}^{a})$
is a geodesic of $M^{4}$ then from (\ref{extradim1}) we must have $%
f^{,a}(y_{o}^{1},...y_{o}^{n})=0$. Let us now calculate the normal
curvatures of $M^{4}$. From (\ref{warpedmetric}) we see that the vectors $%
N_{(a)}=\frac{1}{\sqrt{k_{aa}}}\frac{\partial }{\partial y^{a}}$ (where no
summation is implied over the index $a$) are normal to the submanifold $%
M^{4}.$ Let $\gamma $\ be a curve of $M^{4}$ with tangent vector given by $%
V=(\frac{dx^{\alpha }}{d\lambda },0)$. The normal curvature of $M^{4},$ at a
point $p\in M^{4},$ in the direction of $N_{(a)}$ is given by the inner
product $\Omega _{a}=g(N_{(a)},\frac{DV}{d\lambda })$ at $p$, where $\frac{DV%
}{d\lambda }$ denotes the covariant derivative of $V$ with respect to the
Levi-Civita connection determined by $g$. After computing $\frac{DV}{%
d\lambda }$ from (\ref{warpedmetric}) we can easily show that 
\begin{equation*}
\Omega _{a}=g\left( \frac{1}{\sqrt{k_{aa}}}\frac{\partial }{\partial y^{a}},%
\frac{DV}{d\lambda }\right) =\frac{k_{ab}}{\sqrt{k_{aa}}}\Gamma _{\alpha
\beta }^{b}\dot{x}^{\alpha }\dot{x}^{\beta }=\frac{f_{,a}}{\sqrt{k_{aa}}}%
e^{2f}h_{\alpha \beta }\dot{x}^{\alpha }\dot{x}^{\beta },
\end{equation*}%
where, again, no summation over index $a$ is intended. Thus the coefficients
of the normal curvature $\Omega _{a}=\Omega _{\alpha \beta a}\dot{x}^{\alpha
}\dot{x}^{\beta }$ are given by $\Omega _{\alpha \beta a}=\frac{f_{,a}}{%
\sqrt{k_{aa}}}e^{2f}h_{\alpha \beta }$. \ It is clear that $\Omega _{\alpha
\beta a}=0$ if and only if $f_{,a}=0.$ Therefore, in view of the above we
conclude the following: \textit{timelike or spacelike geodesics can be
confined to the submanifold }$M^{4}$\textit{\ if and only if the normal
curvature of }$M^{4}$\textit{\ vanishes}$.$ It is interesting to ask what
happens in the case of null geodesics. Now since for a curve $\gamma $ to be
a null geodesic of $M^{4}$ $\epsilon $ needs to be zero, equations (\ref%
{spacetime}) and (\ref{extradim1}) imply that $\gamma $ would also be a
geodesic of $M$, irrespective of the value of $f_{,a}$, and hence
irrespective of the value of the normal curvatures of $M^{4}$. Thus the
motion of photons in this setting is not constrained by the extrinsic
curvature of the hypersurface.

The discussion of the geodesics in the previous section allows this theorem
to be generalised to cases where the geometry of the ambient space $M$ is
Riemannian and the hypersurface under consideration is replaced by a
submanifold of codimension $n$. This can be readily seen from Eqs. (\ref%
{extradim1}) by noting that the condition for the vanishing of the extrinsic
(normal) curvatures in this case is given by $f_{,a}=0$ which ensures that
the 4D part of the geodesic Eqs. (\ref{extradim1}) is geodesic.

The above results give the conditions for the geodesics of the higher
dimensional space $M$ to be confined to the co-dimension $n$ hypersurface,
but do not give any information concerning the stability of such a
confinement. We shall consider this question in the following sections,
using a phase space analysis. This generalises a recent result of the
authors concerning the embeddings of hypersurfaces with codimension one \cite%
{DRST}.


\section{Motion in the extra dimensions: an analysis of the phase space}

Defining $\frac{dy^{a}}{d\lambda }=z^{a}$ allows the equations (\ref%
{extradim1})\ to be expressed as an autonomous dynamical system given by 
\begin{align}
\frac{dy^{a}}{d\lambda }& =z^{a}  \label{dynamical} \\
\frac{dz^{a}}{d\lambda }& =P^{a}(z,y),  \notag
\end{align}%
where $P^{a}(z,y)=-$ $f^{,a}(\epsilon +k_{bc}z^{b}z^{c})$ $-\Gamma
_{bc}^{a}z^{b}z^{c}$, with $\ \epsilon =0,1$. Writing the equations as a
dynamical system allows the equilibrium points (given here by $\frac{dy^{a}}{%
d\lambda }=0=\frac{dz^{a}}{d\lambda }$) as well as their stabilities to be
studied. These in turn allow a great deal of information to be gained
regarding the types of behaviour allowed by the system (see e.g. \cite{Smale}%
).

\section{Motion of massive particles in the neighbourhood of the submanifold%
\protect\bigskip\ M$^{4}$}

Let us start with the investigation of the motion of massive particles. For
simplicity we shall first consider the case where the metric $k_{ab}$ is
Euclidean, i.e. $k_{ab}=diag(+1,...+1)$. It is then easy to see that the
equilibrium points exist if the simultaneous equations $f_{,a}=0$ have real
roots (which we denote by $y_{o}=(y_{o}^{1},...,y_{o}^{n})$). The
equilibrium points are then given by $(z^{a}=0,y=y_{o})$.

The nature as well as the stability of these equilibrium points can be
obtained by linearising equations (\ref{dynamical}) and studying the
eigenvalues of the corresponding Jacobian matrix about the equilibrium
points. In this case this is a $2n\times 2n$ matrix given by


\begin{equation}
\Omega =%
\begin{bmatrix}
\label{mat} 0_{n\times n} & 1_{n\times n} \\ 
-\left( f_{ab}\right) _{n\times n} & 0_{n\times n}%
\end{bmatrix}%
,
\end{equation}%
\vskip0.1in \noindent where $f_{,ab}=\frac{\partial ^{2}f}{\partial
y^{a}\partial y^{b}}$, and $0_{n\times n}$ and $1_{n\times n}$ are zero and
unit $n\times n$ matrices respectively. It is not difficult to see that the
determinant and the trace of $\Omega $ ($\det \Omega $ and $I$) are given,
respectively, by $\det \Omega =(-1)^{n}\det $ $f_{,a}$ and $I=0$.

For general forms of the warping function the eigenvalues of the system will
satisfy a polynomial of order $2n$ which would be difficult to analyse
analytically. Our primary aim here, however, is see whether the system is in
principle capable of providing confinement of particles in the neighbourhood
of the spacetime $(3+1)$ hypersurface in the co-dimension $n$ setting. Thus
rather than pursuing the general case we shall ask whether there are classes
of special warping functions $f$ for which such confinement is possible.

As an example we shall consider the cases where $f_{ij}$ computed at the
equilibrium points is zero for all $i\neq j$ and real positive numbers for $%
i=j$. In such cases the matrix simplifies and the eigenvalues (all of which
turn out to be pure imaginary) can be readily found: 
\begin{equation}
\lambda =\pm i\sqrt{f_{,aa}},~~~~a=1....2n,
\end{equation}%
where no summation is intended over $a$. This condition would clearly be
satisfied for warping functions of the type 
\begin{equation}
f=\sum_{a}c_{a}(y^{a})^{n},  \label{function}
\end{equation}%
where $n$ is a positive integer.

Writing the system (\ref{dynamical}) with (\ref{mat}) in terms of normal
coordinates, allows it to be expressed in terms of de-coupled harmonic
oscillators: 
\begin{equation}
\frac{d^{2}y^{j}}{dt^{2}}=-\omega _{j}^{2}y^{j},
\end{equation}%
where $\omega _{j}=f_{jj}$ and again no summation is intended over $j$.

Thus in such co-dimension $n$ cases, one can view the motion in extra
dimensions as confined to n-tori. This generalises the case of the centre
equilibrium point that was found in the co-dimension one case recently \cite%
{DRST} and amounts to a \textit{toroidal confinement} of particles in the
neighbourhood of the spacetime hypersurface.

\section{Confinement through an effective potential}

An analysis of the motion in the extra dimensions can also be carried out by
reducing the problem to the motion of a particle subjected to the action of
an \textit{effective potential} $V=V(y)$. For generality in this section we
shall assume that the metric $k_{ab}$ in Eq. (\ref{warpedmetric}) is a
Riemannian (positive-definite) metric. 

The geodesic equations for the extra dimensions part are given by 
\begin{equation}
\frac{d\dot{y}_{a}}{d\lambda }-\frac{1}{2}k_{bc,a}\dot{y}^{b}\dot{y}%
^{c}+f_{,a}e^{2f}h_{\alpha \beta }\frac{dx^{\alpha }}{d\lambda }\frac{%
dx^{\beta }}{d\lambda }=0.  \label{geodesics1}
\end{equation}%
In these coordinates equation (\ref{firstintegral}) becomes 
\begin{equation}
e^{2f}h_{\alpha \beta }\frac{dx^{\alpha }}{d\lambda }\frac{dx^{\beta }}{%
d\lambda }-k_{ab}\frac{dy^{a}}{d\lambda }\frac{dy^{b}}{d\lambda }=\epsilon ,
\label{firstintegral1}
\end{equation}%
which again allows the decoupling of the motion in the extra dimensions with
equations%
\begin{equation}
\frac{d\dot{y}_{a}}{d\lambda }-\frac{1}{2}k_{bc,a}\dot{y}^{b}\dot{y}%
^{c}+f_{,a}\left( \epsilon +k_{bc}\dot{y}^{b}\dot{y}^{c}\right) =0.
\label{extraDmotion1}
\end{equation}%
A first integral of the above equation may be found by multiplying (\ref%
{extraDmotion1}) by the factor $2\dot{y}^{a}e^{2f}$ to give 
\begin{equation*}
2e^{2f}\left( \dot{y}^{a}\frac{d\dot{y}_{a}}{d\lambda }-\frac{1}{2}\dot{y}%
^{a}k_{bc,a}\dot{y}^{b}\dot{y}^{c}\right) +2\dot{y}^{a}e^{2f}f_{,a}\left(
\epsilon +k_{bc}\dot{y}^{b}\dot{y}^{c}\right) =0,
\end{equation*}%
which, in turn, gives 
\begin{equation*}
e^{2f}\frac{d}{d\lambda }\left( k_{bc}\dot{y}^{b}\dot{y}^{c}\right) +\frac{%
de^{2f}}{d\lambda }\left( k_{bc}\dot{y}^{b}\dot{y}^{c}\right) +\epsilon 
\frac{de^{2f}}{d\lambda }=0.
\end{equation*}%
Integrating we obtain 
\begin{equation*}
e^{2f}\left( k_{bc}\dot{y}^{b}\dot{y}^{c}\right) =C-\epsilon e^{2f}.
\end{equation*}%
Now, if we assume that $f(0,...,0)=0$, then it follows that%
\begin{equation*}
C=\dot{y}_{0}^{2}+\epsilon ,
\end{equation*}%
where\textit{\ }$\dot{y}_{0}^{2}\equiv k_{bc}\dot{y}_{0}^{i}\dot{y}_{0}^{j}$%
\textit{\ }is related to the initial kinetic energy corresponding to the
motion in the extra dimensions.\textit{\ }Thus, we have%
\begin{equation}
e^{2f}\left( k_{bc}\dot{y}^{b}\dot{y}^{c}\right) =\dot{y}_{0}^{2}-V\left(
y\right) ,  \label{Vpotential}
\end{equation}%
where\textit{\ }$V\left( y\right) $\textit{\ } can be treated as an
effective \textit{\ potential energy }given by 
\begin{equation*}
V=\epsilon ^{2}(e^{2f}-1).
\end{equation*}%
Given that the metric $k_{ab}$ is positive-definite the motion is not
allowed in the region where $V(y)>\dot{y}_{0}^{2}$. On the other hand, the
particle may be bound to $\Sigma $ in the neighbourhood of $y=0$\ if $y=0$\
is a point of minimum of $V(y).$ We note, however, that in cases where the
embedding of the submanifold $M^{4}$ has codimension greater than one\ we
can have bounded motion without the particle crossing the brane.

As a simple example we consider the case corresponding to an embedding of $%
M^{4}$ with codimension two, that is, $M=M^{4}\times N^{2}$. Let us assume
the geometry of $N^{2}$ is given by the line element 
\begin{equation}
dl^{2}=u^{2}(r)(dr^{2}+r^{2}d\theta ^{2}).  \label{2dmetric}
\end{equation}%
Clearly we can interpret the \textit{radial }coordinate $r$ as related to
the distance between the points of $M$ and the submanifold $M^{4}$, while
the function $u(r)$ gives a measure of the non-Euclideanicity of the metric $%
k_{ab}$. Also since two-dimensional Riemannian manifolds are conformally
flat, any two-dimensional metric $k_{ab}$ can locally be put in the form (%
\ref{2dmetric}), but in general with $u=u(r,\theta )$ \cite{Novikov}. We now
make the assumption, which in this case seems to be rather natural, that the
warping function $f$ is also a function of $r$, that is, $f=f(r)$. With
these assumptions the equation of the motion (\ref{Vpotential}) yields 
\begin{equation*}
e^{2f}u^{2}\left( \dot{r}^{2}+r^{2}\dot{\theta}^{2}\right) =\dot{y}%
_{0}^{2}-V\left( r\right) .
\end{equation*}%
{}From the equation of motion (\ref{extraDmotion1}) for $\theta $, we get
the following constant of the motion:%
\begin{equation*}
u^{2}r^{2}\dot{\theta}=L=const.
\end{equation*}%
In this way, the equation for the radial motion reduces to 
\begin{equation*}
e^{2f}u^{2}\dot{r}^{2}=\dot{y}_{0}^{2}-\left( V\left( r\right) +L^{2}\frac{%
e^{2f}}{u^{2}r^{2}}\right) .
\end{equation*}%
Here $L$ can be thought of as being related to the angular momentum of the
motion of the particle around the submanifold $M^{4}$. Clearly the existence
of bounded states depends upon the behaviour of the effective potential%
\begin{equation*}
V_{eff}=\left( V\left( r\right) +L^{2}\frac{e^{2f}}{u^{2}r^{2}}\right) .
\end{equation*}

We shall end this section by deducing some general properties of the motion.
Let us assume that $f\left( 0\right) =0$ and that $u(0)$ is regular and does
not vanish at the origin, since otherwise the metric $g$ would not be well
defined at the spacetime $M^{4}$. Now if $L\neq 0$ then the dominant term in
the potential $V_{eff}$ in the limit $r\rightarrow 0$ is $L^{2}\frac{e^{2f}}{%
u^{2}r^{2}}\rightarrow \infty $ . We therefore have our first result (which
is physically obvious) thus: if $L\neq 0$, then the particle cannot return
to the submanifold, whether or not its motion corresponds to a bound state.
On the other hand, for $L=0$, $V_{eff}$ acts as confining potential (for any
value of $\dot{y}_{0}^{2}$) if asymptotically (as $r\rightarrow \infty $),
we have $f\left( r\right) \rightarrow \infty $.

\begin{figure}[tbp]
\includegraphics[height=4.0in,width=4.0in]{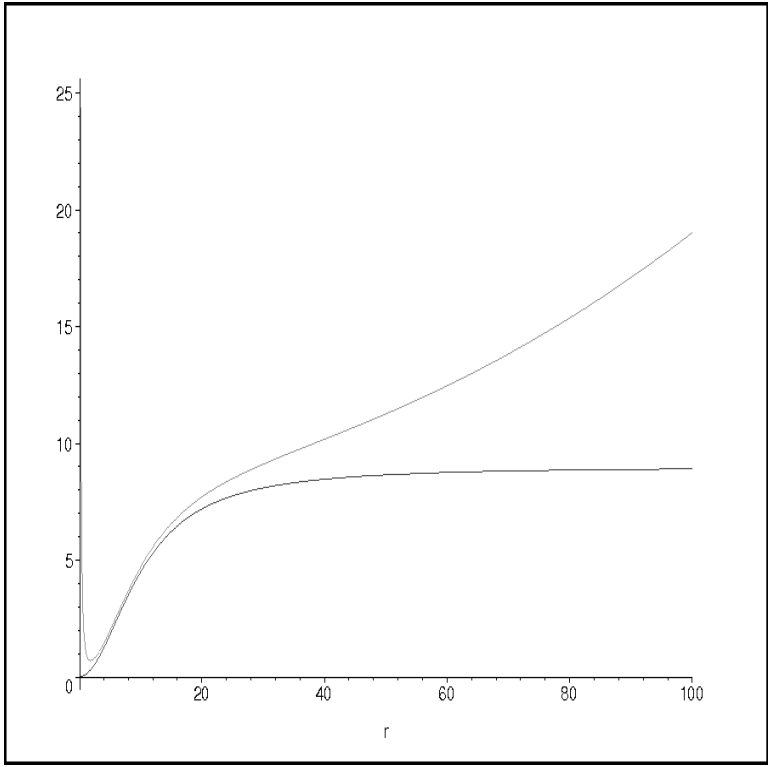}
\caption{Figure showing the behaviour of the effective potential $V_{eff}$
for different values of $L$. For the upper curve (with $L=1$), $%
V_{eff}\rightarrow \infty $ as $r\rightarrow 0$, implying that the particle
cannot cross the submanifold $M^{4}$. For the lower curve (with $L=0$), on
the other hand, the particle is confined to $M^{4}$. The value of the
parameter $c$ was taken to be $c=100$.}
\label{Fig1}
\end{figure}

As a simple application of this result, let us consider a particular case
found in the literature \cite{Gog}, in which the functions $f(r)$ and $u(r)$
are explicitly given by 
\begin{equation*}
e^{2f}=\frac{c^{2}+ar^{2}}{c^{2}+r^{2}}
\end{equation*}%
\begin{equation*}
u^{2}=\frac{c^{4}}{(c^{2}+r^{2})^{2}},
\end{equation*}%
where $a$ and $c$ are real constants. In this case the effective potential
is given by 
\begin{equation*}
V_{eff}=\left[ \epsilon ^{2}\left( \frac{c^{2}+ar^{2}}{c^{2}+r^{2}}\right) -1%
\right] +L^{2}\frac{(c^{2}+ar^{2})(c^{2}+r^{2})}{c^{4}r^{2}}
\end{equation*}%
The behaviour of $V_{eff}$ is displayed in the Fig. \ref{Fig1}, for two
different values of the parameter $L$. The upper curve (with $L=1$)
corresponds to the case where $V_{eff}\rightarrow \infty $ as $r\rightarrow 0
$ and hence the particle cannot cross the submanifold $M^{4}$, despite the
fact that it might be in a bound state. The lower curve (with $L=0$), on the
other hand, corresponds to the case where the particle is confined to $M^{4}$%
. 

\bigskip 

\bigskip 

As can be seen, $V_{eff}$ will act as a confining potential if $a>1$ and $%
L=0 $. In such cases, there exists a limiting value of the kinetic energy
above which the particle will escape to infinity.

\section{Conclusions}

In this paper we have examined some aspects of the motion of massive
particles and photons in a $(3+1+n)$-dimensional warped product spaces.
Spaces of this type, where the codimension of the embedding is one or two,
have received a great deal of attention over the recent years mainly in
connection with the so-called braneworld scenarios. Our treatment has been
geometrical and classical in nature.

We have derived the conditions under which timelike and spacelike geodesics
in the full space $M$ coincide with those on the codimension $n$
hypersurface. We also have shown that the motion of photons does not depend
on the extrinsic curvature. 
Employing the splitting that naturally occurs in such spaces between the
motion in the hypersurface and the remaining dimensions, and using plane
analysis, further allows the stability of such a confinement to be also
studied. Using this approach, we have found a novel form of
quasi-confinement (namely toroidal confinement) which is neutrally stable.
The importance of such confinements is that they are due purely to the
classical gravitational effects, without requiring the presence of brane
type confinement mechanisms.

Finally in connection with our results regarding the conditions that need to
be satisfied by a warped product space in order to ensure the confinement
and stability of geodesic motions, we would like to mention a recent work 
\cite{seahra}, in which the author considers an analogous question. However,
although this analysis is quite general, the author restricts himself to the
case of codimension one embedding. Both approaches would essentially lead to
same results when the warped space is taken to be five-dimensional. \newline

\section*{Acknowledgement}

F. Dahia and C. Romero would like to thank CNPq-FAPESQ (PRONEX) for
financial support.



\end{document}